\documentstyle[prl,preprint,aps]{revtex}

\begin{document}
\draft
\title{New magnetic coherence effect in superconducting 
$\rm{La_{2-x}Sr_{x}CuO_{4}}$}
\author{T.E. Mason$^{1),2)}$, A. Schr\"{o}der$^{2)}$, 
G. Aeppli$^{2),3)}$, H.A. Mook$^{4)}$, and S.M. Hayden$^{5)}$}
\address{$^{1)}$Department of Physics, University of Toronto,
Toronto, M5S 1A7, Canada\\
$^{2)}$Department of Solid State Physics, Ris\o~National Laboratory, 
4000 Roskilde, Denmark\\ 
$^{3)}$NEC Research, 4 Independence Way, Princeton, NJ 08540., U.S.A.\\
$^{4)}$Solid State Division, Oak Ridge National Laboratory, Oak Ridge,
TN 37831, U.S.A.\\
$^{5)}$H.H. Wills Physics Laboratory, University of Bristol, Bristol BS8 1TL,
United Kingdom} 
\date{Physical Review Letters {\bf 77}, (1996), 1604}
\maketitle
\begin{abstract}
We have used inelastic 
neutron scattering to examine the magnetic fluctuations at
intermediate frequencies  in the simplest high temperature superconductor, 
$\rm{La_{2-x}Sr_{x}CuO_{4}}$. The suppression of the low energy magnetic 
response in the superconducting
state is accompanied by an increase in the response at higher
energies.  Just above a threshold energy of $\sim 7$ meV 
there is additional scattering present below T$_{\rm c}$ 
which is characterised by an 
extraordinarily long coherence length, in excess of $50 {\rm \AA}$.
\end{abstract} 
\pacs{PACS numbers: 74.70.Vy, 25.40.F}

Spin pairing is responsible for dramatic reductions in the magnetic 
response of metals as they enter the superconducting state. 
Over the last few decades nuclear resonance\cite{hebel57} and neutron 
scattering measurements\cite{shull66,mason92} have
provided convincing evidence for this suppression, thereby validating
one of the key features of the BCS wavefunction:
the electron spins are hidden because they belong to singlet 
pairs\cite{bardeen57}. Of course, once the pair binding energy
is exceeded, the bare electron spins should become visible again. The 
manner in which this occurs depends on the coherence of
excitations from the superconducting ground state, as well as the
symmetry of the superconducting order parameter.  We have used inelastic 
neutron scattering to examine the magnetic fluctuations at
intermediate frequencies  in the simplest high temperature superconductor, 
La$_{2-x}$Sr$_{x}$CuO$_{4}$. A key conclusion is that 
the suppression of the low energy magnetic response in the superconducting
state is accompanied by an increase in the response at higher
energies.  What is more extraordinary is that at energies just above where
superconductivity begins to enhance magnetic scattering, the enhancement
is very sharp in momentum space.  Indeed, the enhancement is 
resolution limited, corresponding to a coherence length in excess of all
other magnetic, superconducting, and electronic length scales, with the
exception of the superconducting phase coherence length, for
La$_{1.86}$Sr$_{0.14}$CuO$_{4}$.

We performed our experiment with the TAS VI cold neutron spectrometer at
Ris\o~National Laboratory in Denmark using the 
La$_{1.86}$Sr$_{0.14}$CuO$_{4}$
crystals, with $T_{c} = 35$ K, and spectrometer configurations
employed in previous experiments\cite{mason92,mason93}.  The scattered neutron
intensity is proportional to the two spin correlation function which in
turn is proportional to the imaginary part of the
wavevector ({\bf Q})- and energy ($\hbar\omega$)- dependent magnetic
response function multiplied by a thermal population factor,
$(n(\omega)+1)\chi''(${\bf Q},$\omega)$ where
$(n(\omega)+1)=(1-e^{-\hbar\omega/k_{B}T})^{-1}$. Throughout this paper, 
we use the square lattice notation employed by theorists to locate 
wavevectors in the reciprocal space for the CuO$_{2}$ planes
(nearest neighbor Cu-Cu separation=a$_{o}$=3.8 \AA$^{-1}$) of 
La$_{2-x}$Sr$_{x}$CuO$_{4}$.  For La$_{1.86}$Sr$_{0.14}$CuO$_{4}$, 
$\chi''({\bf Q},\omega)$ peaks
at four incommensurate positions, denoted {\bf Q}$_{\delta}$ in Fig. 1 a), 
near the magnetic ordering vector, $(\pi,\pi)$, of the insulating and 
antiferromagnetic parent, La$_{2}$CuO$_{4}$\cite{vaknin87}.  
These correspond to the four
locations equivalent to $(\pi,\pi)+\delta(\pi,0)$ with $\delta=0.245$.
Fig. 2 shows scans along the solid line in Fig. 1 a) through two of the peaks 
for two energy transfers ($\hbar\omega$). It also shows scans along the dashed 
line in Fig. 1 a), which are representative of the temperature-dependent and
weakly {\bf Q}-dependent background. For 
$\hbar\omega = 6.1$ meV,  the peak intensities clearly decrease on passing 
from 
T = 35 K = T$_{c}$ to 5 K, while for $\hbar\omega = 9$ meV, they actually 
increase. The former effect is due
to the usual depletion of the electron-hole pair continuum and has been amply
documented over the last four years for various samples of 
La$_{2-x}$Sr$_{x}$CuO$_{4}$\cite{mason92,mason93,thurston92}. 
The latter effect is new, and being confined to a region very close to the
normal state peak positions, actually corresponds to a dramatic
sharpening of the incommensurate fluctuations in the superconducting state.
In general, one would expect the low frequency suppression of intensity 
to be accompanied by an enhancement at higher frequencies due to 
conservation of spin.
Obviously, an increase in signal between 35 K and 5 K could be associated 
with continuation of an evolution in the normal state
intensity as well as with the onset of superconductivity. To check which
hypothesis is correct, we have measured the peak intensities as a function
of temperature for several $\hbar\omega$. Fig. 3 shows the results. 
For the lowest frequency, 6.1 meV, we see the well-known reversal of the 
growth in the normal state signal at T$_{c}$.  
In contrast, for both $\hbar\omega = 9$ meV and
15 meV, the intensities undergo negligible evolution in the normal state, while
there is a clear increase which  begins on cooling below T$_{c}$. We conclude 
that the rise in incommensurate peak intensity between T$_{c}$ and 5K is due
to superconductivity.  For these higher $\hbar\omega$ the temperature 
dependence of the thermal factor is quite small and is in the wrong direction
to produce an increase in the intensity at low temperatures.
Previous work at these energies did not have sufficient sensitivity to
detect the increase at T$_c$ above 7 meV \cite{mason92,mason93}.

Fig. 4 gives an overview of how the magnetic signal changes on going
into  the superconducting state in the form of a plot, as a function of
both $\bf{Q}$ and $\hbar\omega$, of the differences between the 5 K and 35 K 
data. The important
features are: (i) that the difference crosses zero at 
$\hbar\omega \simeq 7$meV, (ii) that it does not appear to return to zero in 
the range of
frequencies (to 15 meV) accessed above 7.5 meV, (iii) that at 9 meV, the 
difference is a sharp resolution-limited peak centred at the incommensurate 
position (see also the inset in Fig. 2 a)), and (iv) that at  
$\hbar\omega = 15$ meV, the difference appears 
somewhat broader and shifted to seemingly smaller incommensurability. Indeed, 
fitting Gaussians to the difference spectra, we find that the
peak at 15 meV has shifted by $0.04 \pm 0.01$ \AA$^{-1}$ towards the centre of
the scan along the solid line in Fig. 1a) while it is centred at the 
incommensurate
wavevector that characterises the normal state at 9 meV.  The peak in the 
difference spectrum is also broader at 15 meV, with a full width at half 
maximum of $0.10 \pm 0.02$ \AA$^{-1}$ compared to a resolution-limited 
$0.04 \pm 0.01$ \AA$^{-1}$ at 9 meV.  This width indicates a coherence length 
certainly in excess of 50 \AA~and most probably larger than 
100 \AA~($\sim 26$a$_{o}$) for the additional 9 meV excitations 
in the superconducting state.  In addition, this is well 
beyond the 25
\AA~coherence length characterising the incommensurate magnetic fluctuations
above T$_{c}$ at 9 meV\cite{mason92,aeppli96}, as well as the estimated
20-30 \AA~superconducting pair radius for 
La$_{1.86}$Sr$_{0.14}$CuO$_{4}$\cite{ginsberg89}.  The inset to Fig. 2 a) 
shows the
resolution limited Gaussian response (solid line) together with the 
cross section expected for the 25 \AA~length scale of the normal state
using the lineshape which best describes the normal state response 
(dashed line)
\cite{mason92,mason93,aeppli96}.  Clearly the resolution limited peak provides
a better description of the data (indeed models with coherence lengths
of 100 and 1000 \AA~provide equally good fits to the data).  This length 
scale is associated with 2.2 THz spin fluctuations and should not be 
confused with the thermodynamic correlation length.

It is possible to obtain a qualitative understanding of our observations 
by considering Fig. 1 b) and c). In b) we show the Fermi 
surface most likely responsible for the incommensurate peaks in the magnetic
response of La$_{2-x}$Sr$_{x}$CuO$_{4}$\cite{littlewood93}. 
Inelastic scattering occurs when neutrons create electron-hole
pairs: the arrows correspond to the best nesting vector (i.e. the translation 
which brings the largest portion of the Fermi surface into good alignment 
with another parallel portion) and concomitantly the neutron wavevector 
transfer whereby electrons
are removed most efficiently from immediately below the Fermi surface to 
immediately above in the normal state. Superconductivity eliminates these  
low frequency electron-hole pair excitations, as illustrated in c). However, 
c) also reveals that excitations with momenta precisely equal to the nesting 
vector, {\bf Q}$_{\delta}$,
are allowed for an energy transfer equal to the pair binding energy, $2\Delta$.
Indeed,
at this energy, the only excitations allowed have precisely the wavevector
2{\bf k}$_{f}$, implying that the electron-hole pair excitation peak which one
would find for fixed neutron energy transfer, as in Fig. 2,  should be narrower
in the superconducting than the normal state, where the momentum phase space
for possible energy conserving transitions is larger. Of course, as 
$\hbar\omega$
increases, even for the superconducting state, more phase space becomes
available, and there should be a broadening in the enhanced electron-hole
pair response function, as we see in our
experiment.  

It is remarkable that phase space arguments of the type contained in 
Fig. 1 c) account for our data. Indeed, our samples are not optimally 
doped, display considerable pair-breaking in the superconducting state, 
probably have a 
Fermi surface with sufficient curvature to broaden the peak beyond what 
Fig. 1 c) suggests, and are known to host phonons as well as electron-electron 
interactions which would broaden all sharp momentum space features at finite 
frequencies or temperatures. Because each phenomenon just enumerated
contributes to the width of the intensity enhancement seen at 9 meV below 
T$_{c}$, we must conclude that each phenomenon is separately characterised by 
a length scale in excess of 50 \AA. Thus, the effect we have discovered cannot
have anything to do with the disorder which leads to other potentially 
suboptimal properties (e.g. a T$_{c}$ of 35 K instead of 38 K) of our sample. 
Furthermore,  it appears that the peak arises from a unique spanning vector 
across the Fermi surface, implying that its curvature together with the 
{\bf Q}-dependent gap function
are so arranged that the one-dimensional phase space portrait,
shown in Fig. 1 c),  is somehow 
adequate.  Finally, the (electron-electron and electron-phonon) interaction -
derived mean free path, whose inverse contributes directly to the broadening 
in momentum space,  is indistinguishable from infinity in the superconducting 
state.   This result, obtained directly from scans in momentum space,  
has been long conjectured in order to explain anomalies in 
various other 
spectroscopies of the superconducting state\cite{kuroda90}.  Because phonons
do not disappear below T$_{c}$, the scattering of high-frequency 
quasiparticles from phonons should also not change appreciably on passing 
through T$_{c}$.  Because the surplus intensity in the superconducting state is
at least a factor of two narrower (after correction for resolution effects)
than the underlying incommensurate peaks in the normal state,  it is thus 
improbable that electron-phonon scattering plays much of a role even in the 
normal state.

No calculation or previous experiment have anticipated the new phenomenon which
we have discovered, namely a new length scale, indistinguishable from 
infinity, 
characterising a superconductor. Even so, it is worth putting our results in 
the context of other work on $\chi''({\rm {\bf Q}},\omega)$
of high-T$_{c}$ materials. In particular, a 
41 meV resonance appears in the superconducting phase of the bilayer compound 
YBa$_{2}$Cu$_{3}$O$_{7}$, and has spawned 
considerable theoretical activity.  Although there are similarities to what
we have observed in that there is a pronounced enhancement of the response
at finite frequency near the zone boundary there are significant
differences.  The 41 meV feature YBa$_{2}$Cu$_{3}$O$_{7}$ is centered at the 
commensurate wavevector 
$(\pi,\pi)$, is much broader in momentum space (the corresponding coherence 
length is 3-6 \AA), is a narrow peak 
(full-width at half maximum $< 2.5$ meV) in energy\cite{rossat93}
rather than the threshold phenomenon observed in 
La$_{1.86}$Sr$_{0.14}$CuO$_{4}$ where the superconducting to
normal state difference does not recover to zero from 
$\hbar\omega=7$ meV until at least 15 meV.  In addition the 41 meV feature in
YBa$_{2}$Cu$_{3}$O$_{7}$ shows a clear c-axis modulation which indicates it
is related to the CuO bilayers, one of the defining characteristics
of that compound.  Even so, some of the theories for YBa$_{2}$Cu$_{3}$O$_{7}$
also have implications for single-layer La$_{1.86}$Sr$_{0.14}$CuO$_{4}$. 
Demler and Zhang propose a sharp collective excitation at $(\pi,\pi)$
\cite{demler95} as a feature of all cuprate superconductors. 
Given that the enhancement which we have discovered is neither 
sharp in frequency nor centered at $(\pi,\pi)$, this is not the origin of the 
incommensurate signal enhancement which we have discovered for 
La$_{1.86}$Sr$_{0.14}$CuO$_{4}$ to be sharp in momentum space rather than in 
energy. Other approaches combine the band theory of metals displaying 
antiferromagnetic correlations, but not order, with a d-wave pairing state and
give rise to features resembling both the 41 meV peak
in YBa$_{2}$Cu$_{3}$O$_{7}$ and the enhanced scattering we observe in
La$_{1.86}$Sr$_{0.14}$CuO$_{4}$\cite{bulut93,tanamoto93,maki94,onufrieva94,lavagna94,bulut95,zha96}. 
Analogous calculations\cite{liu95,mazin95} for s-wave pairing can mimic the
41 meV resonance for the bilayer materials, but do not yield the enhancement
discovered in the present work for the single layer compound.  
In spite of their success in anticipating an enhanced high-$\omega$ response
in the superconducting state of La$_{2-x}$Sr$_{x}$CuO$_{4}$, the available
calculations miss crucial features of the data, most notably 
the sharpness in {\bf Q} of the enhanced $\chi''$.
They also predict a {\bf Q}-space anisotropy below $\hbar\omega \leq 4$ mev)
in $\chi''$ which, in spite of great experimental effort, has yet to be 
observed in La$_{1.86}$Sr$_{0.14}$CuO$_{4}$\cite{mason93,thurston92}. 

Finally we note that there has recently been great activity on 
real-space domain
pictures yielding incommensurate peaks\cite{tranquada95}.  
It would be interesting to find out whether the strong
coupling effects naturally included in such pictures could account for our 
data.

In conclusion, we have measured the effect of superconductivity on the 
magnetic response near $(\pi,\pi)$ and at intermediate frequencies
(6 meV$ < \hbar\omega <$ 15 meV) in La$_{1.86}$Sr$_{0.14}$CuO$_{4}$.
As the total moment sum rule and the singlet nature of
superconductivity suggest, there is a crossing from negative to positive in 
the difference between magnetic spectra in the superconducting and 
non-superconducting state as a function of energy. However, 
for $\hbar\omega$ just above the zero 
crossing, the spectral weight added by superconductivity is extraordinarily 
sharp, implying that a new length scale indistinguishable from infinity 
characterises the superconductivity in this simple and not especially clean 
high-T$_{{\rm c}}$ material.

We thank K.N. Clausen, K. Bechgaard and J.K. Kjems for their hospitality at 
Ris\o~ and Y. Zha, P. Coleman, C.M. Varma, P. Littlewood, L.P. Gor'kov,
V. Emery, N. Bulut and D. Scalapino for very helpful discussions. This work 
was supported by 
N.S.E.R.C., C.I.A.R., N.A.T.O. the U.S. DOE and the U.K. E.P.S.R.C..

\begin{figure}
\caption[maps]{a) The region of reciprocal space probed in this experiment
showing the trajectories used to measure the signal (solid line) and background
(dashed line). b) Schematic Fermi surface for La$_{1.86}$Sr$_{0.14}$CuO$_{4}$ 
showing the near nesting condition for the spanning wavevector 
{\bf Q}$_{\delta}$.  c) Simple one-dimensional illustration of how the 
possible spin-flip transitions contributing to $\chi''({\bf Q},\omega)$, 
with {\bf Q} spanning the Fermi surface, change upon the development of 
superconductivity.}
\label{map3}
\end{figure}

\begin{figure}
\caption[qscan]{Intensity of the incommensurate magnetic response measured
along the solid line in Fig. 1 a) for 
$\hbar\omega = 9$ and 6.1
meV above (35 K) and below (4.6 K) T$_{c}$.  The background (measured along the
dashed line in Fig. 1 a)) is shown for the same energies and temperatures.  The
data have been normalised to a constant incident monitor, the actual counting
times used in this experiment varied between 25 and 210 minutes per point.  The
error bars reflect the actual counting statistics.  The inset shows the
difference in the intensity measured at the two temperatures for 9 meV.  The
lines are fits described in the text.}
\label{qscan}
\end{figure}

\begin{figure}
\caption[tscan]{Temperature dependence of the incommensurate magnetic response 
at the ${\bf Q}_{\delta}$ position for 6.1, 9, and 15 meV energy transfer. The 
background measured at the cross (for which {\bf Q}=$|${\bf Q}$_{\delta}|$) in
Fig. 1a) has been subtracted from the raw count rates.  For
energies below 7 meV the magnetic response is suppressed below T$_{c}$ while for
higher energies it is enhanced.  The lines are guides to the eye.  The open 
symbols in the bottom panel show the effect of correcting the 6.1 meV data for 
the thermal factor in the cross-section.  For higher energies this 
correction is much smaller and would further enhance the observed increase in
intensity below T$_{c}$.}
\label{tscan}
\end{figure}

\begin{figure}
\caption[eqdiff]{Perspective plot of the difference in the scattered 
intensity in
the ($\hbar\omega$=E,{\bf Q}) plane between the superconducting and 
normal state.  The
momentum dependence is expressed as a function of the displacement in
\AA$^{-1}$ from the incommensurate position, {\bf Q}$_{\delta}$.
The data shown corresponds to half of the
scan of Fig. 2.  The measured intensity difference
(I(4.6 K) - I(35 K)) is plotted together with 
the gaussians which best describe the energy and {\bf Q} dependence observed
(see text).}
\label{eqdiff}
\end{figure}

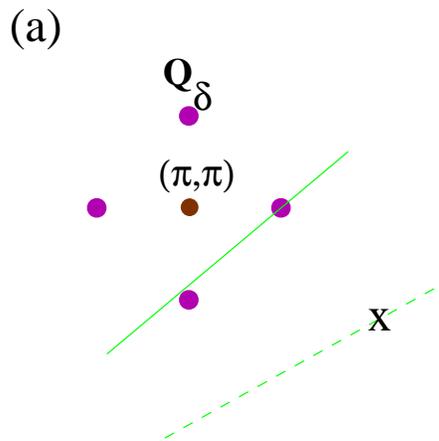
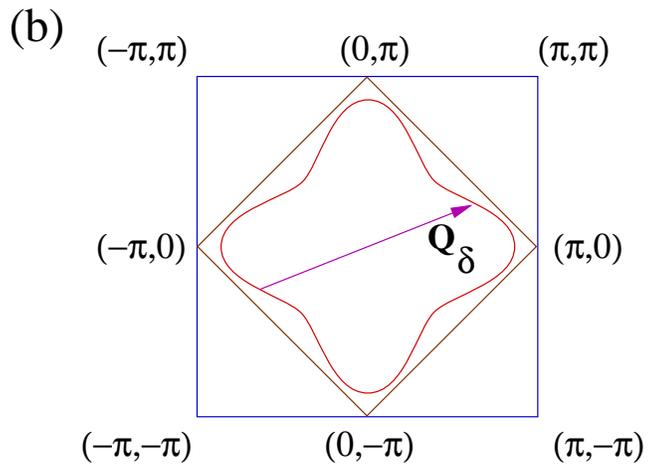
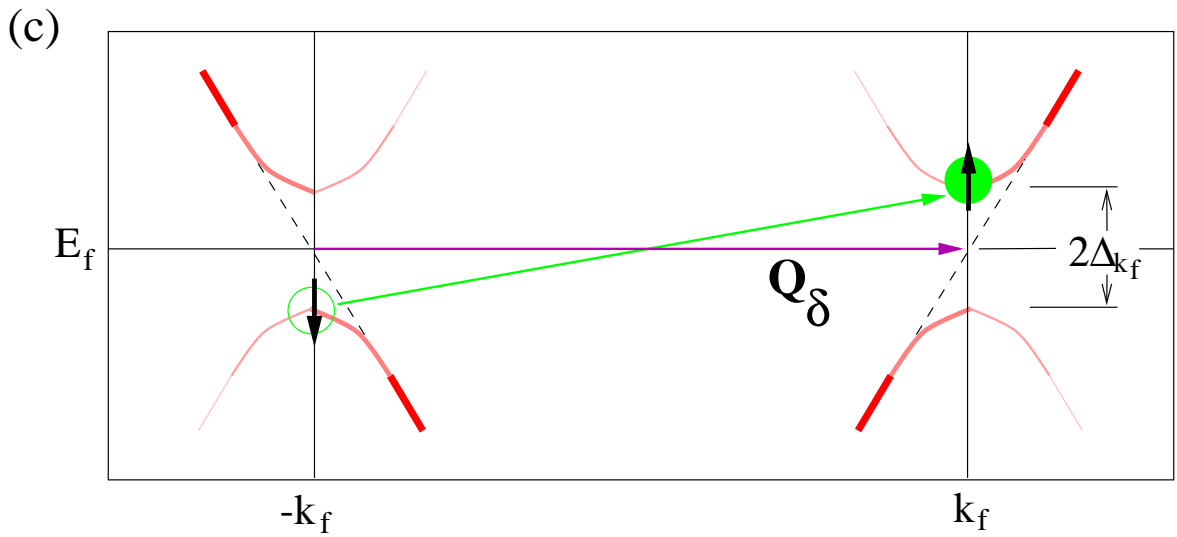
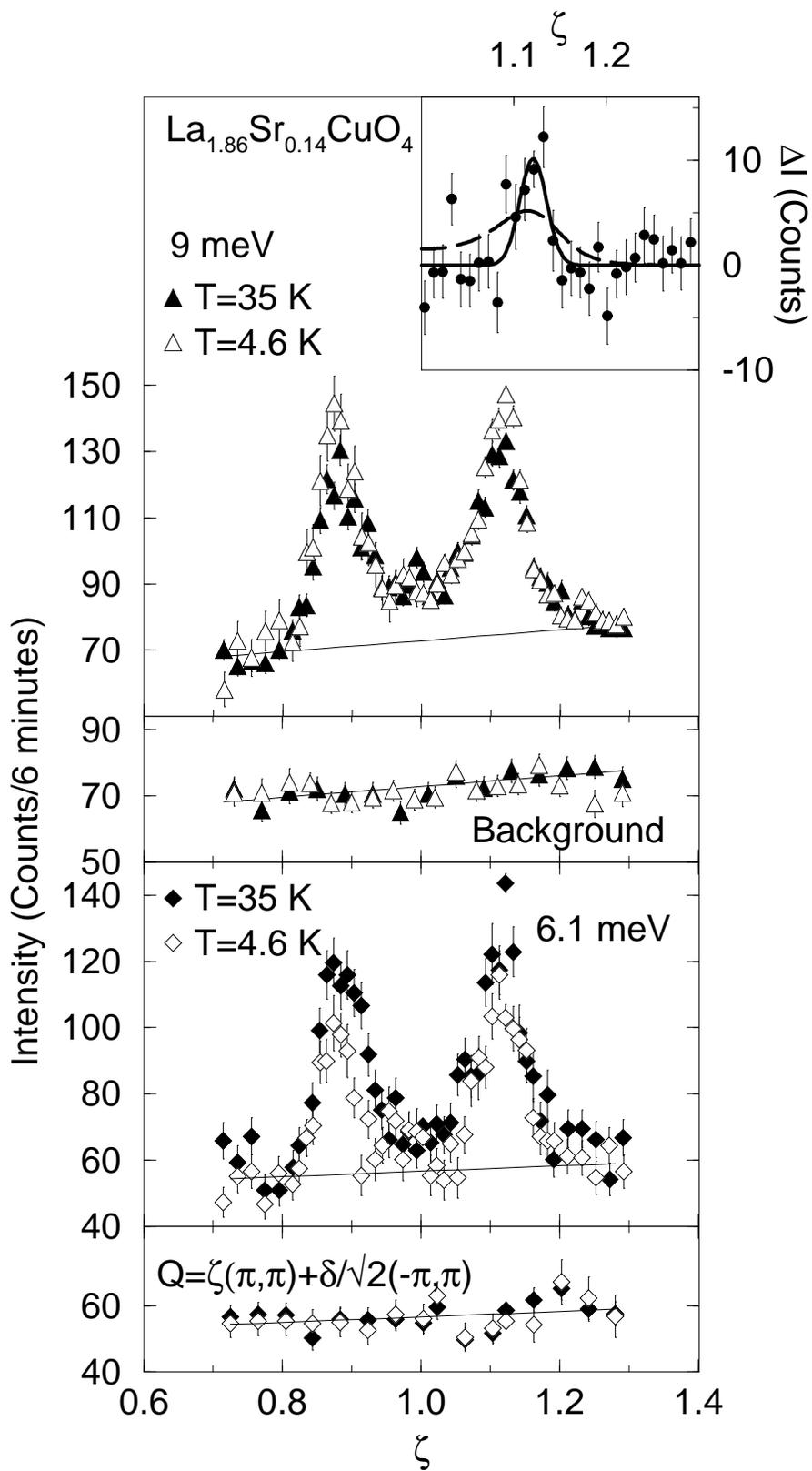
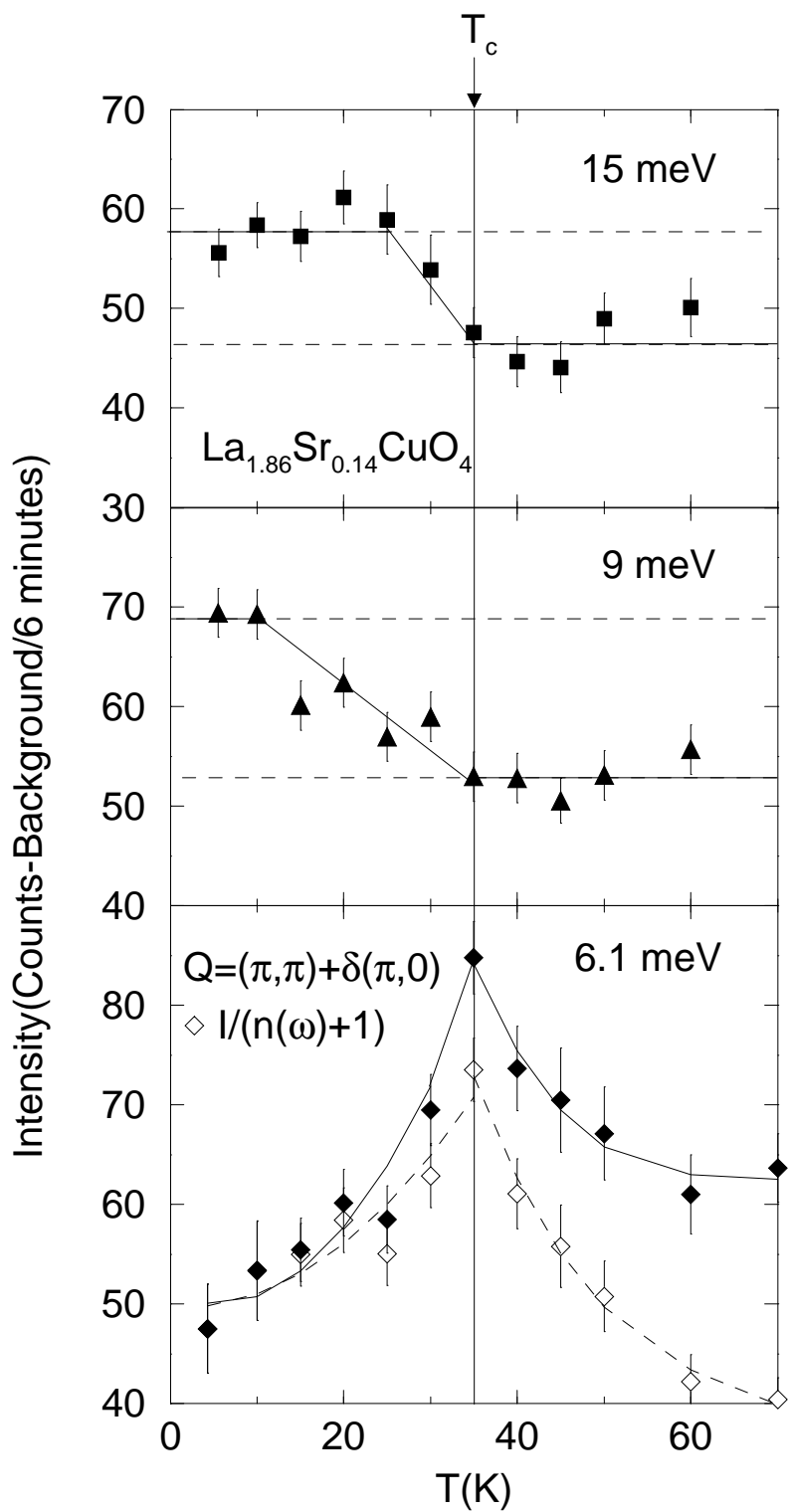
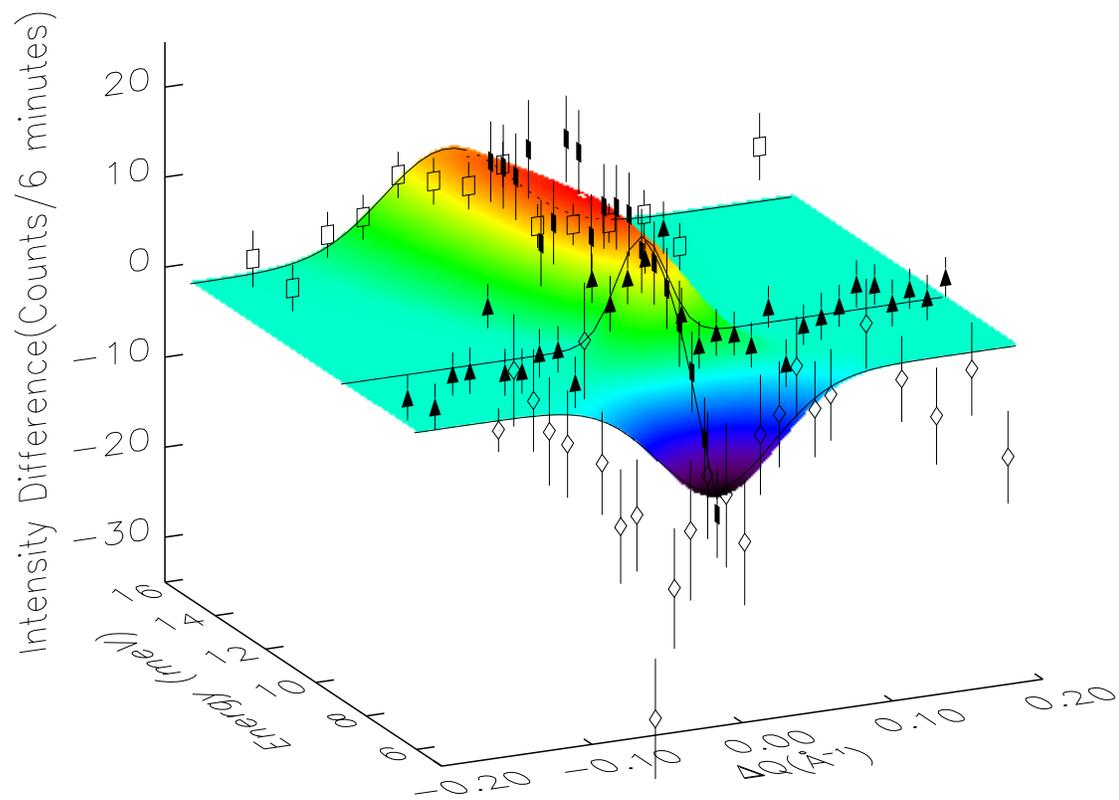
\end{document}